# Non-Intrusive Energy Disaggregation Using Non-negative Matrix Factorization with Sum-to-k Constraint

Alireza Rahimpour, *Student Member, IEEE,* Hairong Qi, *Senior Member, IEEE*, David Fugate, Teja Kuruganti

*Abstract*—Energy disaggregation or Non-Intrusive Load Monitoring (NILM) addresses the issue of extracting device-level energy consumption information by monitoring the aggregated signal at one single measurement point without installing meters on each individual device. Energy disaggregation can be formulated as a source separation problem where the aggregated signal is expressed as linear combination of basis vectors in a matrix factorization framework. In this paper an approach based on *Sum-to-k constrained Non-negative Matrix Factorization* (S2K-NMF) is proposed. By imposing the sum-to-k constraint and the non-negative constraint, S2K-NMF is able to effectively extract perceptually meaningful sources from complex mixtures. The strength of the proposed algorithm is demonstrated through two sets of experiments: Energy disaggregation in a residential smart home; and HVAC components energy monitoring in an industrial building testbed maintained at the Oak Ridge National Laboratory (ORNL). Extensive experimental results demonstrate the superior performance of S2K-NMF as compared to state-of-the-art decomposition-based disaggregation algorithms. The source code and our collected data (HVORUT) for studying NILM for HVAC units can be found at: https://bitbucket.org/aicip/non-intrusive-load-monitoring.

## I. INTRODUCTION

ENERGY disaggregation or Non-Intrusive Load Monitoring (NILM) is the task of decomposing the whole-home energy signal and reporting the operation of individual electrical loads using only measurements of the aggregated signal at the utility service point of a house. In fact, information about individual appliances is more useful to consumers than total electricity usage. This type of feedback is beneficial from different perspectives. Obviously, energy saving is the most important benefit of NILM, but NILM can also be employed on smart grids to determine the operating behavior of loads that are connected to a large scale power distribution system and thus, it would be possible to detect the events that may occur in power grid just by using measurements of a single central sensor. Identifying faulty components in an electrical device by monitoring its power consumption is another possible application of NILM. For instance, if Heating, Ventilating, and Air Conditioning (HVAC) units operate in a degraded or faulted condition, these faults would lead to reduced energy efficiency, reduced comfort, and potential equipment damage. Continuous monitoring of the building equipment is therefore crucial to identify the faults at the early stage and making decisions for repair. As we will describe in this paper, we can monitor the behavior of different components of the HVAC units solely by decomposing the energy signal of the whole building and without installing sensors on individual HVAC components.

### A. Related Works

Recently the NILM has gained major attention in the power system community [1]–[11]. The approaches used for Non-Intrusive Load Monitoring can be categorized into two major classes. The first group includes methods for *load classification* and the second group comprises techniques for *load disaggregation (decomposition)*.

In methods based on classification, there are some distinctive features that define the appliance signature (steady-state or transient-based) and classification is performed based on a learned model. In fact, in load classification the goal is to classify existing appliances at home to several classes where these classes can be defined based on intended application. For example, we can classify appliances according to different factors such as method of load control (e.g., dimmed electrical loads, shed electrical loads, shifted electrical loads) or amount of power usage (low-power, medium-power, high-power). In this widely studied category which initiated by the work of Hart (1992) [12], several works have been done recently by using latent variable models and taking advantage of appliance feature sets [13], [14]. In [13], a non-intrusive approach to classify electrical appliances based on higher-order statistics (HOS) is proposed. Aiming at reducing the computational cost of the proposed method, Fisher's Discriminant Ration and Genetic Algorithms (GA) were used for selecting a finite set of representative features among those obtained by HOS. Tabatabaei, et al., in [14] examined a multi-label classification algorithm (MLkNN), employing both time-domain and wavelet-domain appliance feature sets. In [15] the load current is decomposed into an active and non-active components and the non-active component of the load current is proposed as a distinctive appliance feature, and compared with the traditional load signature. Nevertheless, after many explorations mainly focused on the appliance signature (e.g., [16]), a robust set

Alireza Rahimpour and Hairong Qi are with the Department of Electrical Engineering and Computer Science, University of Tennessee, Knoxville, TN 37996, USA. Email: {arahimpo, hqi}@utk.edu.

David Fugate and Teja Kuruganti are with Computing and Computational Sciences Division, Oak Ridge National Laboratory, Oak Ridge, TN 37831, USA. Email: {fugatedl, kurugantipv}@ornl.gov.

of features that can effectively describe the appliances have not yet been defined [17]. There are several reasons for this deficiency. One reason is that most of the low power appliances have similar power usage and similar electrical features that lead to the difficulty of their classification. In addition, most of the multi-state appliances, e.g., washing machines, have particular user settings and their power consumption pattern is not unique [18]. Moreover, depending on the features selected to describe the appliance, even the building electrical network can influence the captured signature. Therefore, finding some distinctive appliance signatures is one of the most indispensable issues when the energy disaggregation is seen as a classification problem [19]. A review of existing techniques mainly based on classification can be found in [17], [20], [21].

The second group of NILM methods look at the problem as decomposing a mixture signal into individual appliances signals (based on single-channel source separation) and formulate the task as an optimization problem. [22] is the first work which employed a sparse coding approach for source separation on the energy disaggregation task. We will discuss the details of this method in the next section. The decomposition based approaches rely on constructing a model which includes signatures of different devices as its bases. Linear decomposition of the aggregated signal (via matrix factorization) using bases of this learned model (instead of predefined bases) results in an efficient estimation of the energy consumption of each device. Learning the model from the training samples instead of using some predefined bases such as Fourier or wavelet bases has been shown to produce more accurate results [23]–[28].

There are two key points in using different types of source separation methods. The first one is whether the model is learned in a supervised or unsupervised fashion. Unsupervised learning techniques have been widely used to automatically discover the underlying structure of data. This may serve several purposes, depending on the task at hand [29]–[31]. For instance, in experimental sciences, one may be looking for data representations that automatically exhibit interpretable patterns. Independent component analysis (ICA) is most known unsupervised method for blind source separation [32]. What distinguishes ICA from other decomposition methods is that it looks for components that are both statistically independent, and non-Gaussian. However, in general, ICA cannot identify the actual number of source signals, a uniquely correct ordering of the source signals, nor the proper scaling (including sign) of the source signals. All these, limit the application of this method.

The existing unsupervised methods for load disaggregation include different use of hidden Markov models (e.g., Difference HMM, etc.) [33]–[35]. Unsupervised methods require hand tuning of parameters, which can be challenging for the methods to be generalized in practice. On the other hand, in supervised methods the model can be built based on the labeled training samples. The existing supervised methods for NILM include using sparse coding [22], change detection [36] and dynamic modeling approaches [37]. In the proposed approach in this paper, the model is learned in supervised manner during the training stage of the algorithm.

The second important issue in designing matrix factorization methods is choosing proper constraints for the specific problem at hand. For example, in [38] by adding minimum volume constraint to NMF, the algorithm can successfully decompose the highly mixed pixels in hyper spectral images in unsupervised manner. One important constraint that can be added to matrix factorization problem is sparsity constraint using $L1$ norm. Even though the $L1$ norm was first introduced in geophysics [39], it was popularized in statistics with the Lasso estimator of Tibshirani (1996) [40] and independently in signal processing with the basis pursuit formulation of Chen et al. [41]. Many other constraints depending on the task at hand, can be added to matrix factorization problem (e.g., [42]–[49]). We discuss this topic in more details in Section II.

### B. Contributions

We refer to the proposed approach in this paper as Sum-to-k constrained Non-negative Matrix Factorization (S2K-NMF). The contribution of this paper is three-fold. First, a new regularization term is designed in the form of sum-to-$k$ constraint in matrix factorization framework, which induces grouping effect and reduces the adverse effect of correlation between bases in the dictionary and yields improvement of the decomposition result in the energy disaggregation task.

Second, in contrary to the existing works, in this paper we study the energy disaggregation task for estimating the energy consumption of every single known device in the dataset, regardless of the size or type of the loads. In fact, we will show that the proposed algorithm is able to estimate the profile of the energy consumption of all the existing appliances (at the home or building testbed dataset) using real low-frequency aggregated signal (i.e., for both current and power) without the need of feature extraction from each appliance's signal. Many of disaggregation algorithms use high-frequency data. However, sampling at high-frequency requires expensive hardware and installation of new monitoring devices in a building. Furthermore, some other event-based methods either require extracting features for each device or are only capable of detecting some limited states (e.g., ON or OFF only) of some of the devices.

Third, to the best of our knowledge, this is the first work that successfully demonstrates the disaggregation of the HVAC components in an industrial building environment. In fact, besides solving the energy disaggregation problem in the smart home environment, we further apply the proposed S2K-NMF in the context of smart industrial building where the aggregated power signal of the whole building is decomposed into the power consumption of different parts of the HVAC unit in a hierarchical fashion.

Through extensive experimental evaluation on two different testbeds (i.e., home and building), we show that the proposed S2K-NMF is robust and versatile to handle disaggregation problems at different scales and with different levels of complexity. The remainder of this paper is organized as follows. Section II elaborates on the problem formulation, as well as the proposed S2K-NMF approach. Section III describes the experimental results and the final section concludes the paper and discusses future works.

## II. METHODOLOGY

In source separation problems, several signals have been mixed together into a mixture signal and the objective is to recover the original component signals from the mixture signal. The classical example of a source separation problem is the cocktail party problem, where a number of people are talking simultaneously in a room (for example, at a cocktail party), and a listener is trying to follow one of the discussions. Matrix Factorization (MF) is an effective approach to deal with the source separation problem, which can approximate a mixture signal through linear combination of some bases. Although, standard MF algorithms do not impose any constraint on these bases, to solve real-world application problems and find well-posed formulations, certain kinds of constraints need to be added. The matrix factorization problem can be generally formulated as:

$$\min_{A} \left\| \bar{X} - DA \right\|_F^2 \quad s.t. \ certain \ constraints \quad (1)$$

where in the context of this paper, each column of $\bar{X} \in R^{m \times d}$ is the aggregated signal obtained from a central point (e.g., central electrical meter at home) and $m$ is the dimension of the signal. We consider each column of $\bar{X}$ to be one day of energy consumption signal and we have $d$ days of testing signal. Each column of the signature matrix $D \in R^{m \times T}$ (i.e., each base), is the energy consumption profile (e.g., current or power) of an individual component (e.g., appliance at home) and $T$ is the number of bases. Also, each row of $A \in R^{T \times d}$ is the activation coefficient for the corresponding base in $D$ and $F$ is the Frobenius norm. In the following, we first briefly describe the decomposition-based disaggregation approach [22] in Section II-A. We then elaborate on the proposed Sum-to-k constrained Non-negative Matrix Factorization (S2K-NMF) method in Section II-B. Although on the surface S2K-NMF carries some similarities with another group-based decomposition approach, called Elastic Net, we describe the fundamental difference between these two approaches in Section II-C.

### A. The Sparsity Constraint for Energy Disaggregation

As mentioned in Section I-A, one of the key points for designing a matrix factorization problem is to find an appropriate application-specific constraint. For example, adding sparsity constraint (via $L1$ norm) to the activation coefficient has been a popular approach for signal decomposition in recent years. By adding this constraint to the primary framework (Eq. 1), the optimization problem (called lasso) would be as the following form:

$$\min_{A} \left\| \bar{X} - DA \right\|_F^2 + \beta \left\| A \right\|_1 \quad (2)$$

where $\|A\|_1 = \sum_{j=1}^{T} A_j$ is the $L1$ norm which imposes the constraint that the activation matrix $A$ be sparse (i.e., they contain mostly zero entries) and $\beta$ is the coefficient that controls the level of sparsity. In [50] Marial et al., have answered this question that "why the $L1$ norm induces sparsity?" from analytical, physical and geometrical point of views. Indeed, imposing the sparsity allows us to learn overcomplete representations of the data. It means, there exists more basis functions than the dimensionality of the data in the dictionary (i.e., $D$).

However, using this configuration of data would not be beneficial in energy disaggregation. For instance, in [22] an overcomplete dictionary was build and the dimension of the signal is kept low (e.g., only one sample is used per hour) to guarantee the overcompleteness (i.e., $m << d$), such that the sparsity constraint would be valid. However, in doing so, we are running the higher risk of not capturing the transient signature of each appliance and thus jeopardizing the device identification result. Therefore, having more samples (i.e., higher dimensionality) for each basis, would be more favorable than taking advantage of overcompleteness of the dictionary and imposing the sparsity constraint.

### B. Sum-to-k constrained Non-negative Matrix Factorization

In this section we elaborate on the Sum-to-k constrained Non-negative Matrix Factorization (S2K-NMF) energy disaggregation approach. There are two major constraints in S2K-NMF that are added to Eq. 1. Considering the non-negativity nature of energy signal, one constraint is the non-negative constraint for activation coefficients. Another important constraint is the sum-to-k constraint for activation coefficients. This sum-to-k constraint imposes the "grouping" effect where the basis vectors from the same individual component form a "group".

For example, if we choose 3 days of energy consumption profile for an appliance, say TV, and 2 days of the same for the dishwasher, to form the bases, then $D = [D_1, D_2] = [d_1, d_2, d_3, d_4, d_5]$ where $D_1 = [d_1, d_2, d_3]$ and $D_2 = [d_4, d_5]$ are the two groups describing the signature of the corresponding appliance. Similarly, the activation matrix is grouped using the same structure, i.e., $A = [A_1, A_2]^T = [a_1, a_2, a_3, a_4, a_5]^T$ where $A_1 = [a_1, a_2, a_3]^T$ and $A_2 = [a_4, a_5]^T$. Here, we use $k$ to represent the number of groups, or individual devices for decomposition purpose, such that $D = [D_1, D_2, \ldots, D_k]$ and $A = [A_1, A_2, \ldots, A_k]^T$. Please note that if there is more than one instance of each device in home, they account as separate groups. For instance, in above toy example, if there is a second TV in home, it forms another group and we would have 3 groups (i.e., two TVs and one dishwasher, $k = 3$).

The effectiveness of this sum-to-k constraint is two-fold. First, the elements (i.e., $a_{ij}$) of the activation matrix for each device (i.e., $A_i$) are the probabilities of that device being represented via some bases of the signature matrix (i.e., $D$). Therefore, we enforce the summation of all the probabilities for each device to be equal to one, so that we can be confident, that device is being represented by some linear combination of the bases corresponding to that specific device. Correspondingly, summation of the elements of each column of matrix $A$ is equal to $k$ (e.g., number of devices at home). Furthermore, since the algorithm uses the linear combination of the basis vectors to estimate the test signal (instead of choosing only one specific basis vector), the test signal does not have to exactly match one of the training profiles in the training set (in order to achieve a low disaggregation error). Therefore, there is no need to include all the possible usage patterns of an appliance in the training set and the algorithm is able to perform accurately even with low amount of training set.

Second, the sum-to-k constraint eliminates the adverse effect of correlation between bases of different devices due to its grouping structure. In other words, for calculating the activation coefficients for each device (i.e., $A_i$), the algorithm only looks at the bases corresponding to that specific device (i.e., $D_i$) and not all the bases in the signature matrix. Correlation between the bases of the signature matrix is a major problem in matrix factorization-based decomposition approaches. For instance, in energy disaggregation task, basis vectors of different devices are usually highly correlated and it is difficult to approximate the aggregated signal as a linear combination of these highly correlated bases. By introducing the sum-to-k constraint we can successfully address this issue.

By performing matrix factorization, we calculate the activation coefficients $A_i \in R^{n \times d}$ for each device. The following equation illustrates our linear model and corresponding optimization problem before imposing any constraints:

$$\hat{A}_{1:k} = \underset{A_{1:k} \geq 0}{\operatorname{argmin}} \left\| \bar{X} - [D_1, \dots, D_k] \begin{bmatrix} A_1 \\ A_2 \\ \vdots \\ A_k \end{bmatrix} \right\|_F^2 \quad (3)$$

where $\bar{X}$ is the aggregated signal, $A_i (i = 1, \dots, k)$ is the activation matrix for the $i$th device's base matrix ($D_i$) and $F$ is the Frobenius norm. The $\hat{A}_{1:k}$ in Eq. 3 is equivalent to $\hat{A}_i$ for $(i = 1, \dots, k)$.

After calculating the activation coefficient for each device, the estimated signal for the $i$th device would be:

$$\hat{X}_i = D_i \hat{A}_i \quad (4)$$

We can impose the sum-to-k constraint for the $A$ matrix by adding the second term in the following equation:

$$\hat{A}_{1:k} = \underset{A_{1:k} \geq 0}{\operatorname{argmin}} \left\| \bar{X} - [D_1 \dots D_k] \begin{bmatrix} A_1 \\ A_2 \\ \vdots \\ A_k \end{bmatrix} \right\|_F^2 + \beta \left\| U - QA \right\|_F^2 \quad (5)$$

We refer to Eq. 5 as Sum-to-k constrained Non-negative Matrix Factorization (S2K-NMF). In this equation, $U \in R^{k \times d}$ is a matrix with all its entries equal to one, and $\beta$ is a small weight. $Q$ is the matrix including 1 and 0 elements that we would define in some way that it forces the summation of activation coefficients for each device in matrix $A$ to be equal to one. Let us consider the same toy example as described in the second paragraph of Sec. II-B, where we have $k = 2$ appliances (TV and dishwasher) with 3 days of training data for TV and 2 days of training data for dishwasher (i.e., the total number of columns of $D$ is $T = 5$). Consequently, the signature matrix D would be of size $m \times 5$, where $m$ is the number of samples in each day (i.e., dimensionality of each column). By defining $Q \in R^{k \times T}$ as the following matrix, we impose the sum-to-k (i.e., $k = 2$) constraint for the coefficients of the activation matrix $A$:

$$Q = \begin{bmatrix} 111 & 00 \\ 000 & 11 \end{bmatrix} \quad (6)$$

For estimating the energy consumption of TV, the algorithm only uses the linear combination of $d_1$, $d_2$ and $d_3$ and ignores $d_4$ and $d_5$. The reason is that the proposed constraint (second term in Eq. 5), forces the $a_1 + a_2 + a_3 = 1$ and $a_4 = a_5 = 0$, which is why the first row of $Q$ for this example has the form of [111 00]. Considering the formulation of the constraint in Eq. 5, it imposes: $(1 \times a_1) + (1 \times a_2) + (1 \times a_3) + (0 \times a_4) + (0 \times a_5) = 1$, which means using the linear combination of basis vectors of the TV, (such that the summation of the activation coefficients is equal to one) and ignoring the basis vectors of the dishwasher, and therefore getting rid of the possible correlation between basis vectors of TV and dishwasher.

We employ matrix augmentation for solving the optimization problem in Eq. 5, which leads to the following optimization problem:

$$\hat{A} = \underset{A \geq 0}{\operatorname{argmin}} \left\| \begin{bmatrix} \bar{X} \\ \beta U \end{bmatrix} - \begin{bmatrix} D \\ \beta Q \end{bmatrix} A \right\|_F^2 \quad (7)$$

where in this equation all the matrices are defined the same as Eq. 5. By employing this matrix augmentation, we can incorporate our constraints into the regular matrix factorization problem. For solving the optimization problem in Eq. 7 we adopt a fast version of the Non-Negative Least Square method (FNNLS) based on active set [51].

### C. S2K-NMF vs. Elastic Net

The goal of this section is to clarify the difference between S2K-NMF and another sparsity-constrained decomposition algorithm that also takes advantage of the grouping effect. As we discussed in Section II-A, the algorithm which uses the $L1$ norm for imposing sparsity is called lasso. The $L1$ norm penalty is suitable for some applications, however when there exist some grouping effect in $D$, lasso fails to perform group selection. In fact, if there is a group of highly correlated variables, then the lasso tends to select one variable from a group and ignore the others. To overcome this limitation, the Elastic Net adds a quadratic part to the penalty as shown in Eq. 8 (i.e., the third term). In [52] Zou and Hastie showed that in real world data, the Elastic Net often outperforms the lasso, while enjoying a similar sparsity of representation.

$$\min_A \left\| \bar{X} - DA \right\|_F^2 + \beta_1 \left\| A \right\|_1 + \beta_2 \left\| A \right\|_2 \quad (8)$$

Despite the name similarities (group-based), Elastic Net is a totally different approach from S2K-NMF and is basically a smooth version of the lasso. Elastic Net allows the correlated bases to be selected into or out of a model together, so that all the members of a particular group are either considered or discarded (this type of grouping structure is not always useful in the energy disaggregation task, where the basis vectors of one device can be highly correlated and that device can be totally discarded by the algorithm).

The grouping structure of S2K-NMF and Elastic Net is fundamentally different. In fact, in the Elastic Net, a "group" is merely a collection of arbitrarily correlated bases, whereas in S2K-NMF, we would define the groups based on the specific task at hand (i.e., the "group" has physical meaning). For comparison purpose, we apply the Elastic Net method for the energy disaggregation task and analyze the results in Section III.

## III. EXPERIMENT AND RESULTS

In this paper we design two different experiments for evaluating our proposed algorithm. The first experiment is disaggregation of the whole home energy to the energy consumption of all the appliances at a residential home. The second experiment is designing a hierarchical scheme for disaggregating the whole building energy signal to the HVAC components signals in an industrial building. Although all the figures only present results of selected cases (e.g., one random test day in 20% of the data), they are provided only to give readers better illustration on the result of load disaggregation.

### A. Datasets

In order to implement an accurate and realistic energy disaggregation task, the dataset being used is better to contain all the devices signals and the real aggregated signal in high sampling frequency (i.e., more than one sample per second) for a long enough period of time (i.e., at least one year). In addition, current signal is a more suitable measure for device identification than real power signal. In our experiments we use both power and current signal, but we prioritize on current because the hardware cost and installation cost of current transducers are less than those of other measurements. Moreover, power meters provide power measurements but do not typically provide a "power waveform". Thus, the information from a power meter measurement is limited. [53] discusses how using current is more beneficial in energy disaggregation task based on comprehensive experiments on 473, 232 data points over 11 months. They have discussed some issues such as changing the restiveness (R) of the loads due to some factors such as wire gauge and material used.

Unfortunately, most of the available data sets either have low sampling frequency or or just cover a short period of time or do not include real aggregated signal. Furthermore, some of them are not freely available for research purpose. However, it is worth noting that our proposed algorithm does not need high frequency data or huge training set to work accurately.

In this paper we use the *AMPds: A Public Dataset for Load Disaggregation and Eco-Feedback Research* [53] for the first experiment. This dataset has most required features for performing an accurate disaggregation task. *AMPds* at the time of writing this paper, contains one year of data between April 1, 2012 and March 31, 2013, including different types of power signal, current and voltage that includes 11 measurements at a sampling rate of one sample per minute for 21 sub-meters. The rationale behind choosing this data set over others is that first of all, it includes the real aggregated signal for power and current signals and second, it includes the current signal for all the appliances and for long enough time periods of the data as well. The time duration of data is important because we want to capture the diverse behaviors of several devices in different seasons and include them in our model at the training stage. For example, some devices such as heater and cooler systems have totally different usage behavior in different months of the year.

For the second experiment, the data was collected on the Oak Ridge National Laboratory (ORNL) Flexible Research Platform (FRP1) between February $1^{st}$ 2015 to February $21^{st}$ 2015 (i.e., 21 days). FRP1 was constructed to enable research into building envelope materials, construction methods, building equipment, and building instrumentation, control, and fault detection [54]. FRP1 is a $40 \times 60$ foot sheet metal building that was developed as a partnership with the Metal Building Manufacturers Association. FRP1 is heavily instrumented with measurements of outdoor weather, indoor temperature, humidity, air flow, HVAC refrigerant, and power consumption. FRP1 consists of one large HVAC conditioning zone with the ability to operate a Nordyne gas-heat and electric air conditioning system. The Nordyne system has two compressor stages and their associated condenser fans, gas heat, as well as a standard circulation blower motor. This results in five large electrical loads in the Nordyne unit that can be examined for this research. The sampling rate of the data is 2 samples per minute which is a descent sampling rate for the intended application.

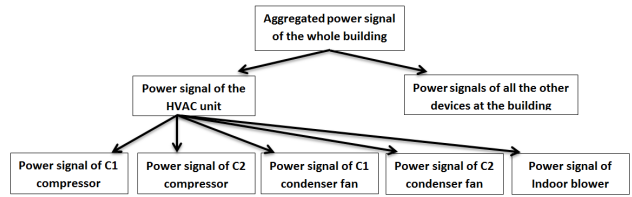

Fig. 1. Hierarchical HVAC energy decomposition

### B. Experimental Design

*1) Residential Home Energy Disaggregation:* In this experiment, the AMPds dataset with sampling rate of 1 sample per minute is used. We take 80% of the data as training set and the rest 20% for the testing using a 10-fold cross-validation. In fact, in the testing stage just the aggregated signal is accessible and the goal is to estimate the energy consumption for all the devices at home. All the 19 appliances and plugs signals at home available in the AMPds data set are shown in first column of Table I. One of the differences between our approach and other works based on sparse coding for disaggregation is that we have omitted the pre-training part. In pre-training [22], first the appliances signal $D_i$ is decomposed to two matrices $B_i$ and $A_i$ for each appliance individually and matrix $B_i$ is used as base for the $i$th device in the next step of the training stage. In this way, some errors will be added due to matrix decomposition, which is not desirable. Instead, in our algorithm the normalized appliance's signal $D_i$ is used as the basis for each device in the dictionary. Experiments show that this leads to increased accuracy and decreased computational cost of the algorithm. For comparison purpose, we correspondingly implement the Discriminative Disaggregation Sparse Coding (DDSC) in [22]. In DDSC implementation, we choose the value of sparsity penalty coefficient equal to 0.01 for imposing enough degree of sparsity. Furthermore, we also implement the Elastic Net algorithm and compare the results ($\beta_2 = 0.001, \beta_1 = 0.01$ in Eq. 8).

*2) HVAC Components Disaggregation:* The goal of the second experiment is to estimate the energy usage profile of the internal components of the HVAC by disaggregating the energy consumption of the whole building. The same as the first experiment, in this experiment 80% of the data is used as the training set and 20% of the data is assigned as the testing set in a 10-fold cross-validation scheme.

Our experiment in this part consists of two major parts: 1-Disaggregation of the power signal of whole building to power signals of all the circuits and devices existing in the building. There are 16 different devices, circuits and plugs in the building: HVAC unit, 480/208 Transformer, lighting circuits: 1, 3, 5, 7, Plug circuits: 1, 3, 5, 7, cord reel circuit, lighting control box, exhaust fan, piping heat trace, exterior lighting (lighting and emergency) and building control circuit. 2-Decomposition of the obtained HVAC power signal from the previous step and estimating the power consumption profile of its components including: two compressors, two condenser fan and one indoor blower. Figure 1 illustrates this hierarchical architecture.

### C. Performance Metrics

For performance evaluation of estimated signals and also to compare different implemented methods in this paper we have exploited different measures: *Root Mean Square Error* (RMSE), *Disaggregation Error* (DE), and Percentage of Contribution in Energy Consumption (PCEC). Moreover, energy usage profile pie graph and plot of estimated signals and their ground truth are other means of comparing the results of different algorithms.

*RMSE* is a frequently used measure of the differences between values estimated by a model or an estimator and the values actually observed. For $d$ days of testing and $m$ samples for each day we have the following formula for RMSE.

$$RMSE(X, \hat{X}) = \sqrt{\frac{\sum_{j=1}^{d} \sum_{i=1}^{m} (X_{ij} - \hat{X}_{ij})^2}{m \times d}} \quad (9)$$

RMSE presents a sample to sample comparison between the estimated signal $\hat{X}$ and the ground truth $X$. *Disaggregation Error* (DE) is another metric which provides a global comparison between the estimated signal and the ground truth. Disaggregation error can be formulated as follows:

$$\sum_{i=1}^{k} \frac{1}{2} \left\| X_i - \hat{X}_i \right\|_2^2 \quad (10)$$

As before, $k$ in Eq. 10 is the number of all present electrical appliances at home. The PCEC is calculated by the amount of energy that the device consumed divided by the whole consumed energy at home in the test day.

### D. Results and Discussion

*1) Residential Home Energy Disaggregation:* The most important aspect which we are looking for in energy disaggregation is estimating the amount of power usage and percentage of contribution of each device in the whole home energy consumption. Table I shows the energy consumption contributions for all the electrical appliances present at home. The second column of this table is the ground truth of PCEC of whole house for each appliance or plug during one day of testing. The last three columns are estimated percentage of contribution of all the appliances using S2K-NMF, Elastic Net and DDSC, respectively. This table is very informative about details of each device's energy consumption and it seems to be more interpretable than other evaluation metrics. Table I illustrates that in a random day chosen for testing (i.e., a random day within 20% of the data as the test set), some appliances are OFF during all day and the S2K-NMF algorithm is the only method which is able to detect this pattern with 100% accuracy (no false positive). For example, clothes washer is OFF during the test day (i.e., Ground truth of *PCEC* is zero) and only S2K-NMF can distinguish that this device does not work at all in that day.

Likewise, the S2K-NMF result does not show any miss detection of a device being OFF unlike what has happened for

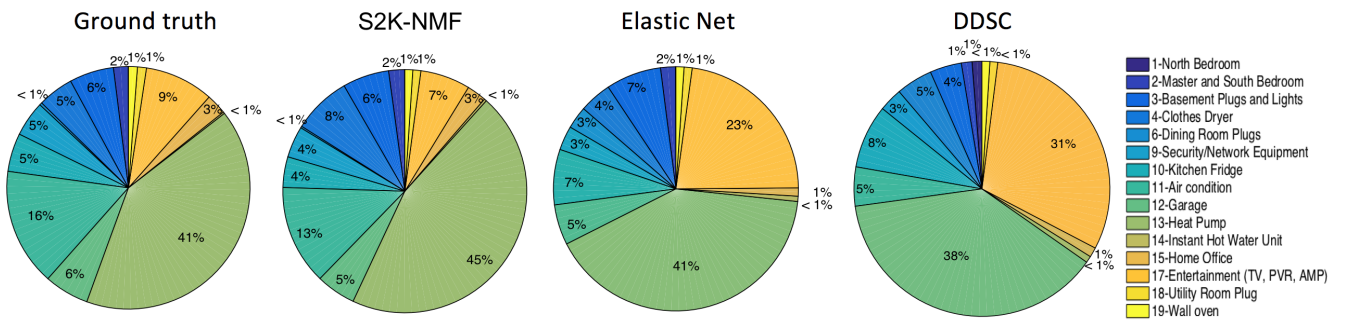

Fig. 2. The pie plots show that S2K-NMF achieves the best result for estimating the energy usage contribution of each device. Best viewed in color.

TABLE I
GROUND TRUTH AND ESTIMATED PERCENTAGE OF CONTRIBUTION IN ENERGY CONSUMPTION (PCEC) BY EACH DEVICE DURING ONE DAY.

| Appliance Type | Ground truth of PCEC% | Estimated PCEC% S2K-NMF | Estimated PCEC%Elastic Net | Estimated PCEC%DDSC [22] |
|---|---|---|---|---|
| North Bedroom | 0.042 | 0.035 | 0.063 | 0 |
| Master and South Bedroom | 2.043 | 2.106 | 2.277 | 1.574 |
| Basement Plugs and Lights | 6.194 | 6.158 | 8.491 | 1.701 |
| Clothes Dryer | 4.999 | 7.741 | 4.974 | 5.035 |
| Clothes Washer | 0 | 0 | 0.171 | 0.176 |
| Dining Room Plugs | 0.355 | 0.220 | 3.448 | 3.825 |
| Dishwasher | 0 | 0 | 0 | 0 |
| Electronics Workbench | 0 | 0 | 0 | 0 |
| Security Equipment | 4.947 | 4.171 | 3.730 | 2.960 |
| Kitchen Fridge | 5.366 | 4.074 | 7.539 | 8.161 |
| A/C Fan and Thermostat | 15.660 | 13.252 | 0 | 0 |
| Garage | 6.249 | 5.266 | 6.217 | 6.242 |
| Heat Pump | 41.080 | 45.336 | 40.829 | 38.007 |
| Instant hot water | 0.233 | 0.319 | 0.802 | 0.832 |
| Home office | 2.847 | 2.591 | 1.274 | 1.550 |
| Outside Plug | 0 | 0 | 0 | 0 |
| Entertainment(TV,PVR,AMP) | 9.605 | 6.617 | 22.690 | 30.618 |
| Utility Room Plug | 1.249 | 1.249 | 1.243 | 0.980 |
| Wall Oven | 1.249 | 1.249 | 1.243 | 0.980 |

TABLE II
AVERAGE OF THE DISAGGREGATION ERROR FOR TRAINING AND TESTING STAGES FOR EXPERIMENT 1 (HOME ENERGY DISAGGREGATION)

| | Training | | | Testing | | |
|---|---|---|---|---|---|---|
| | S2K-NMF | ELASTIC | DDSC | S2K-NMF | ELASTIC | DDSC |
| DE | 0 | 0.092 | 0.171 | 0.880 | 1.671 | 1.989 |

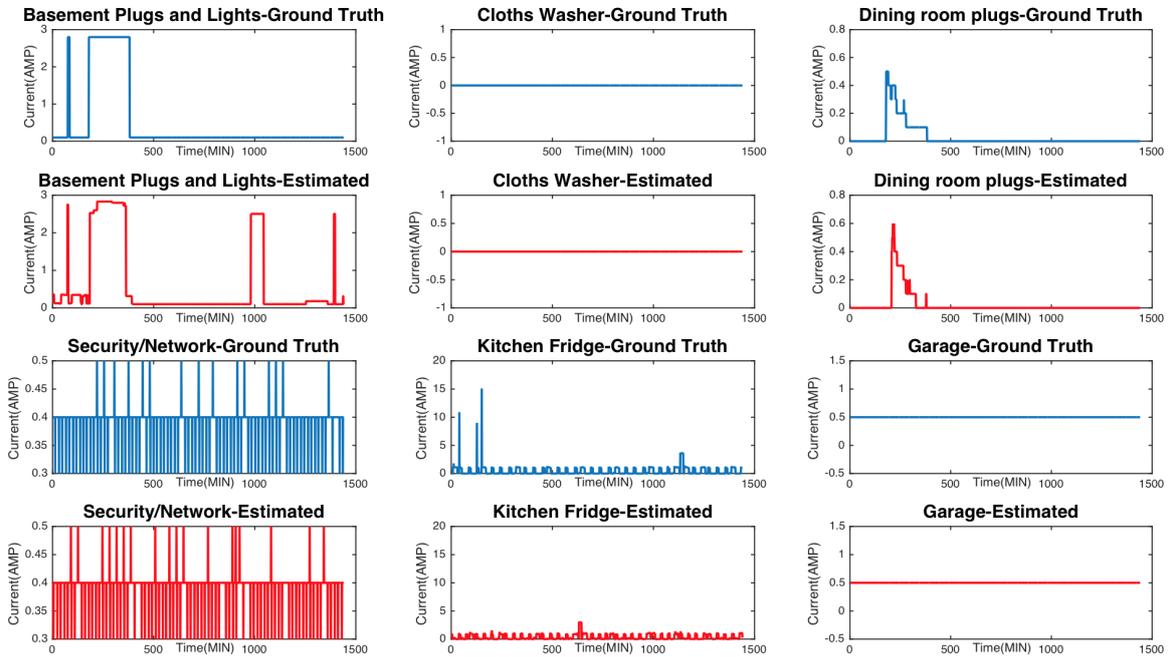

Fig. 3. Ground truth and estimated appliances' signals using the *S2K-NMF* method for one random testing day (1440 minutes). Best viewed in color.

air condition (fan and thermostat) using two other algorithms. From Table I we can also interpret the usage behavior of different devices in one day. For example, major energy is consumed by heat pump (43%), Air condition (Forced Air Furnace: Fan and Thermostat) (16.44%) and Entertainment:

TV, PVR, AMP (9%) as we expect for a typical house. Furthermore, Figure 2 illustrates the pie plot of the contribution of each appliance in energy consumption of the house. Some appliances are OFF during the test day for energy estimation and that is the reason their names are not involved in the pie chart. From Figure 2 and Table I we observe that S2K-NMF achieves the best disaggregation performance by having the closest PCEC% values to the ground truth. The Elastic Net method ranks the second and the DDSC method performs the worst among the three. Table II shows results of comparing performance of different methods in the training and testing stage. In this table DE is the disaggregation error as is defined by Eq. 10. In order to calculate the performance evaluation metrics of this table, 80% of the data including 292 days of energy consumption is used for the training and 20% of the data including 73 days is used for the testing in a 10-fold cross-validation framework. There are total number of 19 appliances at home and we use one day of the energy consumption as one column in the dictionary. Hence, the dictionary size would be: $19 \times 292 = 5548$ for this experiment. Results in Table II illuminate the fact that minimum disaggregation error for both training and testing parts belongs to the S2K-NMF algorithm. Especially, the noticeable numbers at training part of the results are presenting perfect signal decomposition using S2K-NMF which is distinctively better than results of other works in this part. Moreover, from entries of Table I and Table II one can conclude that adding the $L2$ norm constraint (i.e., third term in Eq. 8) to the activation matrix in the Elastic Net method (plus a slight level of sparsity), improves the grouping effect and leads to better results in comparison to non-negative sparse coding approach. To better understand the accuracy of the proposed algorithm, Figure 3 illustrates the appliances' current signal and their estimation using S2K-NMF. These signals have been calculated using our linear model in the testing stage, by having aggregated signal as our only measurement. This figure shows that there are still some inevitable errors in estimating the appliances current signal. Nevertheless, for some devices, the algorithm can estimate the energy without any error. For example, the constant signal of the garage (signal in the 4th row and 3rd column of Figure 3) can be estimated by S2K-NMF quite accurately.

TABLE III
AVERAGE OF THE DISAGGREGATION ERROR (DE) FOR DIFFERENT COMBINATIONS OF TRAINING SET AND TESTING SET USING S2K-NMF IN
EXPERIMENT 1

| Train - Test | 80% - 20% | 70% - 30% | 60% - 40% | 60% - 50% | 40% - 50% | 30% - 70% | 20% - 80% |
|---|---|---|---|---|---|---|---|
| DE | 0.880 | 0.925 | 0.889 | 0.967 | 0.950 | 0.964 | 0.953 |

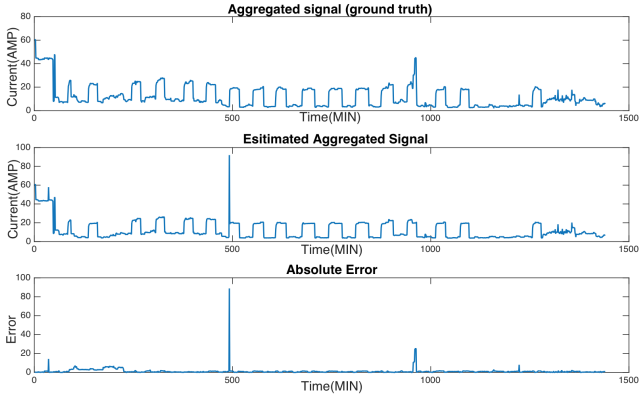

Fig. 5. Ground truth (top figure) and estimated aggregated signal (middle figure) and absolute difference between them (bottom figure) for the residential home in one random testing day (1440 minutes) using the S2K-NMF algorithm.

Figure 5 also shows the aggregated signal and its estimation and their absolute difference using the S2K-NMF method. It is worth noting that in our experiments, if we use summation of the signals instead of real aggregated signal (similar to the other works, e.g., [22]) we would obtain higher accuracy in the disaggregation task but, as it was discussed before, this assumption is not realistic. Furthermore, in order to show the generalization of the proposed method, we study the effect of using different splits of the data for training and testing set on disaggregation error. We create case studies with different numbers of training vs. testing samples: 80% - 20% (i.e., 292 days of training vs. 73 days of testing), 70% - 30%, 60% - 40%, 50% - 50%, 40% - 60%, 30% - 70%, and 20% - 80%, where x% - y% indicates that x% of the entire data is used for training and y% is assigned for testing). For each split, a 10-fold cross validation is conducted, and the reported disaggregation error (DE) takes the average of the cross validations. The first row of Table III illustrates the amount of data for training set and testing set, respectively. This table shows that the lowest disaggregation error is achieved when using 80% of the data as training and 20% as the testing. Also, we observe from Table III that the performance degradation from 80% - 20% to 20% - 80% (i.e., using only 20% of the data as training set) is not large.

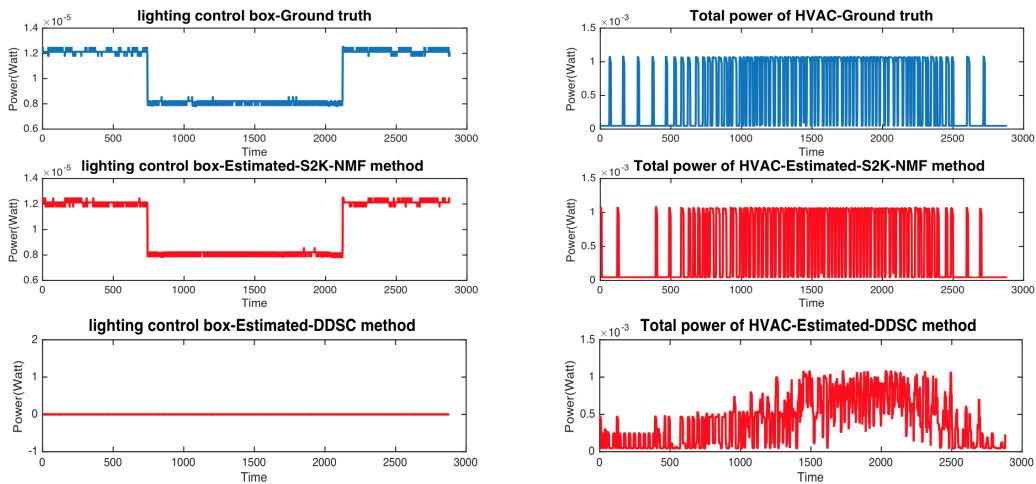

Fig. 4. Building power disaggregation. Two top figures: Ground truth of the power signals of the lighting control box and the HVAC unit. Middle figures: The estimated power consumption profile of the lighting control box and the HVAC unit via the S2K-NMF method during one random day. Two bottom figures: Estimated power signals using the DDSC method [22].

TABLE IV
Average of the Disaggregation error for training and testing stages for experiment 2 (Building Energy Disaggregation)

| | Training | | | Testing | | |
|---|---|---|---|---|---|---|
| | S2K-NMF | ELASTIC | DDSC | S2K-NMF | ELASTIC | DDSC |
| DE | 0 | 0.051 | 0.135 | 0.795 | 1.387 | 1.650 |

TABLE V
Average RMSE for estimation of power consumption profile of different components of HVAC for different methods(C2 condenser fan and Indoor blower are OFF during the test days).

| HVAC Components | Training RMSE-S2K-NMF | Testing RMSE-S2K-NMF | Testing RMSE-Elastic Net | Testing RMSE-DDSC |
|---|---|---|---|---|
| C1 compressor | 0 | 0.004 | 0.086 | 0.946 |
| C1 condenser fan | 0 | 0.001 | 0.020 | 0.147 |
| C2 compressor | 0 | 0.002 | 0.062 | 0.449 |
| C2 condenser fan | 0 | 0 | 0.013 | 0.041 |
| Indoor blower | 0 | 0 | 0.013 | 0.041 |

TABLE VI
Average of the Disaggregation Error (DE) for different combinations of training set and testing set using S2K-NMF in Experiment 2. $DE_1$: DE for disaggregation of the building power signal to the appliances, $DE_2$: DE for disaggregation of the HVAC power signal to its components

| Train - Test | 80% - 20% | 70% - 30% | 60% - 40% | 50% - 50% | 40% - 60% | 30% - 70% |
|---|---|---|---|---|---|---|
| $DE_1$ | 0.795 | 0.896 | 0.890 | 0.970 | 0.978 | 0.973 |
| $DE_2$ | 0.835 | 0.849 | 0.888 | 0.952 | 0.961 | 0.976 |

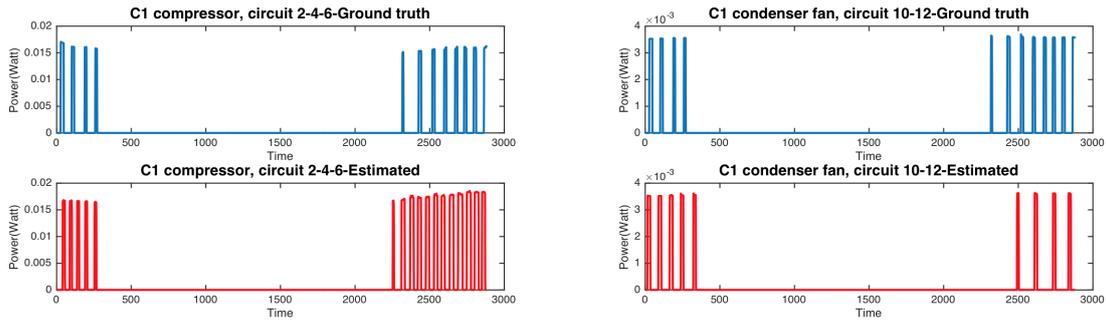

Fig. 6. Estimated power consumption profile of compressor (left) and condenser fan (right) of the HVAC unit for one day using the S2K-NMF. Top figures: Ground truth, Bottom figures: Estimated signals.

*2) HVAC Components Disaggregation:* Table IV shows the disaggregation error for decomposing the power signal of the whole building to the power signal of the HVAC unit and all the other devices in the building. It can be observed from the table that the S2K-NMF achieves the smallest DE in both training and testing. Moreover, Figure 4 demonstrates the estimated power consumption profiles as well as the ground truth signal of the HVAC and lighting control box at the building during one random test day. It is clear that the power estimation using S2K-NMF (middle plots) is more accurate than the results of DDSC (bottom plots). Table V illustrates the RMSE for power disaggregation of the HVAC to its component (C1 compressor, C1 condenser fan, C2 compressor, C2 condenser fan and Indoor blower) using the S2K-NMF, Elastic Net and DDSC. This table demonstrates the superior performance in estimation of the power consumption of each part of the HVAC using the S2K-NMF. It is worth noting that for calculating the RMSE values in Table V, the same 10-fold cross-validation framework described in the first experiment is employed. By looking into the HVAC data, we noticed that the C2 condenser fan and indoor blower have been OFF during the test day. Table V (last two rows) shows that the proposed S2K-NMF method can detect this "not operating" event with 100% accuracy. One potential application of using this monitoring property is for fault detection. In other words, if any of the internal components of the HVAC stop working, we can detect this event using the decomposition of the aggregated signal of the whole building.

We use the same setups of splitting the training vs. testing dataset as the first experiment (i.e., smart home) on the smart building testbed with a 10-fold cross-validation and calculate the Disaggregation Error (DE). The result is shown in Table VI, where the training - testing data vary from 80% - 20% (i.e., 17 days of training vs. 4 days of testing), 70% - 30%, 60% - 40%, all the way to 30% - 70%. The $DE_1$ in Table VI is the disaggregation error for the first layer of the disaggregation (i.e., disaggregation of the energy signal of the building to the appliances), and $DE_2$ is the disaggregation error for the second layer of the disaggregation (i.e., disaggregation of the HVAC power signal to its components). The results in Table VI show that the larger the training data, the more accurate the estimations are. However, the performance degradation from 80% - 20% to 30% - 70% is not large, which confirms the ability of the proposed S2K-NMF to work with small training sets. Moreover, Figure 6 shows the estimated power consumption profile of one of the compressors and condenser fans of the HVAC unit using the S2K-NMF approach.

## IV. CONCLUSION AND FUTURE WORKS

In this research we addressed an effective solution to the energy disaggregation problem by exploiting Sum-to-k constrained Non-negative Matrix Factorization algorithm. As the results showed, adding the sum-to-k constraint to the activation coefficient matrix led to considerable improvement in estimation accuracy. Indeed, the proposed approach for imposing this constraint is applicable to many other tasks which can be formulated as matrix factorization problem.

One essential limitation in energy disaggregation research is lack of a comprehensive data set. Our proposed algorithm works well with low sampling rate data, but as we discussed in this paper, in general, data with high sampling rate can better reveal the transient characteristics of each individual electrical appliances and leads to more effective device identification. Therefore, collecting a data set for energy disaggregation application which includes real aggregated signal and current signal of all the devices at home with high sampling rate that cover a duration of at least one year of the data would be valuable for energy disaggregation research in the future. Towards this goal, we have started to collect the first energy disaggregation dataset (with one year duration) for HVAC components monitoring in an industrial building located at ORNL. We observed that by exploiting the proposed scheme in this paper, we were able to monitor the signals of the internal components of the HVAC system. We can assume that changes in electric power use, length of run time, and degree of cycling under similar outdoor weather conditions for these internal components can be used to infer degradation in the actual efficiency and capacity of the HVAC, when evaluated over appropriate time periods. As a result, these changes can potentially be used for detection of performance degradation and fault detection of this equipment and therefore, as input to guiding maintenance decisions which is a line for our future research. Another good application of the proposed S2K-NMF algorithm is demand response, so that utility would know which device is "ON (demand) and thus turn it off when needed, or vice versa.

The decomposition method introduced in this paper uses supervised learning and it requires to have the history of the power (or current) signal of each device in order to train the model. One direction for future work would be energy disaggregation in unsupervised manner (i.e., estimation of each device's signal without using the history of its energy consumption). The proposed algorithm for electrical energy disaggregation is also applicable to gas and water consumption decomposition and by incorporating them (and also taking advantage of their inter-dependency), we can form a comprehensive energy monitoring system in one package, which can be a line for future works as well.

## ACKNOWLEDGMENT

This work was supported by the Department of Energy's Building Technology Office as part of the 2014 Building Energy Efficiency Frontiers and Incubator Technologies initiative with UT-Battelle and Oak Ridge National Laboratory, in part by the Engineering Research Center Program of the National Science Foundation and the Department of Energy under NSF Award Number EEC-1041877. Also, in part by the CURENT Industry Partnership Program and NSF-CNS-1239478. The authors acknowledge that this research was enabled by the use of instrumentation data from the ORNL Flexible Research Platforms and the instrumentation data support of ORNL staff Anthony Gehl, Joshua New, and Jibo Sanyal.

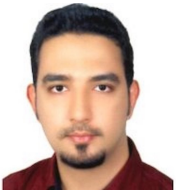

**Alireza Rahimpour** received the B.S. degree in Electrical Engineering, Electronics from Shiraz University, Shiraz, Iran, in 2009 and the M.S. degree from Amirkabir University of Technology (Tehran Polytechnic) Tehran, Iran in 2012. He joined the Department of Biomedical Engineering, Azad University Tehran, Iran as a Lecturer in 2012. He has been with the Electrical Engineering and Computer Science Department at University of Tennessee at Knoxville, TN, USA as PhD candidate since 2013. He received the Electrical Engineering and Computer Science Department Excellence Fellowship and Outstanding Graduate Teaching Assistant Award from the University of Tennessee, Knoxville in 2013 and 2016 respectively. He also received the University of Tennessee Chancellor's award and Outstanding Graduate Research Assistant award in 2017 and 2018. His current research interests include Machine learning, Computer Vision, Artificial intelligence, Signal and Image processing and Non-Intrusive Load Monitoring (NILM). Since 2013, he also has been collaborating as a researcher and mentor with Center for Ultra-Wide-Area Resilient Electric Energy Transmission Networks (CURENT), that is a National Science Foundation Engineering Research Center jointly supported by NSF (National Science Foundation) and the DoE (Department of Energy). He is an IEEE member.

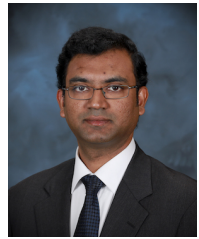

**Teja Kuruganti** is Senior R&D staff member at Oak Ridge National Laboratory (ORNL), where he worked since 2003. He currently leads ORNL activities, as sub-program manager, in developing novel sensors and controls for improving energy efficiency of buildings and novel techniques for enabling grid-responsive building loads. His research interests include wireless sensor networks, communications systems, control systems, and sensor development. He won an R&D 100 award in 2012 for co-developing electromagnetic wave propagation simulation engine for harsh environments. He earned MS and PhD degrees in Electrical Engineering from University of Tennessee, Knoxville and BE in electronics and communication engineering from Osmania University. He has authored over 85 peer-reviewed publications in journals, conferences, and book chapters. He is a member of IEEE and ISA. He is currently the Director of ISA Test and Measurement Division.

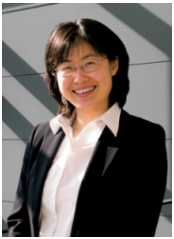

**Hairong Qi** (S'97-M'00-SM'05) received the B.S. and M.S. degrees in computer science from Northern JiaoTong University, Beijing, China in 1992 and 1995 respectively, and the Ph.D. degree in computer engineering from North Carolina State University, Raleigh, in 1999. She is currently the Gonzalez Family Professor with the Department of Electrical Engineering and Computer Science at the University of Tennessee, Knoxville. Her research interests are in advanced imaging and collaborative processing, hyperspectral image analysis, and bioinformatics. Dr. Qi is the recipient of the NSF CAREER Award. She also received the Best Paper Awards at the 18th International Conference on Pattern Recognition (ICPR) in 2006, the 3rd ACM/IEEE International Conference on Distributed Smart Cameras (ICDSC) in 2009, and IEEE Workshop on Hyperspectral Image and Signal Processing: Evolution in Remote Sensor (WHISPERS) in 2015. She is awarded the Highest Impact Paper from the IEEE Geoscience and Remote Sensing Society in 2012.

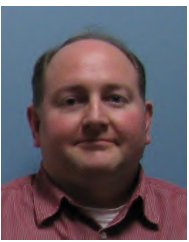

**David Fugate** (S'90-M'92) was born in Rogersville, TN, in 1970. He received the B.S. and M.S. degrees in electrical engineering from Tennessee Technological University of Cookeville, TN in 1994 and 1999. David has worked primarily in the field of electrical engineering and control systems for Rockwell Automation, Engineering Design; Manufacturing Services, Kimberly-Clark, and Pratt& Whitney before joining the Sensors and Controls Research team at Oak Ridge National Laboratory (ORNL). Davids experiences include conceptual development, system design, construction, installation, fielding, and support of sensors, measurement systems, and control systems for manufacturing, research, and aerospace applications. David has significant experience with automation, electrical system design, robotics, induction heating, original equipment manufacturing, process heating, web converting and processing, process and system modeling, control algorithm design, fault detection, embedded systems, software development and testing, and mission-critical systems. During his employment at Pratt; Pratt & Whitney, David was a member of the Control System Design and Integration Team for the F135 propulsion system. David led the control system requirements, design, algorithms, system modeling, and testing for various sensing and actuation subsystems including a novel electro-thermal engine ice protection system. At Oak Ridge National Laboratory, David has participated or led projects related to sensing, control, and wireless communications for nuclear research, building research, and transactive energy.